\documentstyle[12pt]{article}
\newlength{\extraspace}
\newlength{\extraspaces}
\newcommand{\be}{\begin{equation}
\addtolength{\abovedisplayskip}{\extraspaces}
\addtolength{\belowdisplayskip}{\extraspaces}
\addtolength{\abovedisplayshortskip}{\extraspace}
\addtolength{\belowdisplayshortskip}{\extraspace}}
\newcommand{\ee}{\end{equation}}

\newcommand{\ba}{\begin{eqnarray}
\addtolength{\abovedisplayskip}{\extraspaces}
\addtolength{\belowdisplayskip}{\extraspaces}
\addtolength{\abovedisplayshortskip}{\extraspace}
\addtolength{\belowdisplayshortskip}{\extraspace}}
\newcommand{\ea}{\end{eqnarray}}

\newcommand{\newsection}[1]{
\vspace{15mm}
\pagebreak[3]
\addtocounter{section}{1}
\setcounter{equation}{0}
\setcounter{subsection}{0}
\setcounter{footnote}{0}
\begin{center}
{\large \thesection. #1}
\end{center}
\nopagebreak
\medskip
\nopagebreak}

\newcommand{\newsubsection}[1]{
\vspace{1cm}
\pagebreak[3]

\addtocounter{subsection}{1}
\noindent{ \sc \thesubsection. #1}
\nopagebreak
\vspace{2mm}
\nopagebreak}

\newcommand{\hf}{{\textstyle{1\over 2}}}

\newcommand{\is}{\! & \! = \! & \!}
\newcommand{\nonu}{\nonumber \\[1.5mm]}

\def\fr#1#2{{#1\over #2}}
\def\du#1{\tilde #1}

\def\m{\mu}
\def\n{\nu}
\def\r{\rho}
\def\s{\sigma}

\def\f{\phi}
\def\F{\Phi}
\def\FB{{\bar \F}}
\def\ff{\varphi}

\def\th{\theta}
\def\td{\du\th}
\def\o{\omega}
\def\a{\alpha}

\def\d{\delta}
\def\e{\epsilon}
\def\l{\lambda}
\def\L{\Lambda}
\def\k{{\bf k}}         
\def\LB{{\bar \L}}
\def\pa{\partial}
\def\pab{\bar\pa}

\def\ie{{\it i.e.},\ }
\def\vs{{\it vs.}\ }
\def\ind{{1\over 2\pi}\int\!\! d^2\! z\,}
\def\bd{\bar D}
\def\bk{{\bar K}}
\def\ins{{1\over 2\pi}\int\!\! d^2\! z\, D_+D_-\bd_+\bd_-\,}
\def\A{A}
\def\AB{{\bar A}}
\def\J{J}
\def\JB{{\bar J}}
\def\xd{\du x}
\def\q{\, , \qquad}
\def\sq{\, , \quad}
\def\R{\relax{\rm I\kern-.18em R}}
\def\[{\Bigl[}
\def\]{\Bigr]}
\def\gr{G^R_a}
\def\gl{G^L_a}
\def\j{\pa\th_L+B\pa\th_R+\fr12\gl\pa x^a}
\def\jb{\pab\th_R+B\pab\th_L+\fr12\gr\pab x^a}
\def\ra{\rightarrow}
\def\re{{\rm Re}}
\def\im{{\rm Im}}

\catcode`@=11           
\dimendef\dimen@=0
\def\ialign{\everycr{}\tabskip\z@skip\halign} 
\def\m@th{\mathsurround=\z@}
\def\openup{\afterassignment\@penup\dimen@=}
\def\@penup{\advance\lineskip\dimen@
  \advance\baselineskip\dimen@ \advance\lineskiplimit\dimen@}
\def\eqalign#1{\null\,\vcenter{\openup\jot\m@th
  \ialign{\strut\hfil$\displaystyle{##}$&$\displaystyle{{}##}$\hfil
\crcr#1\crcr}}\,}
\catcode`@=12           

\def\,{\kern .16667em}  

\begin{document}
\addtolength{\baselineskip}{.7mm}
\thispagestyle{empty}
\begin{flushright}
{\sc ITP-SB}-91-53\\ {\sc IASSNS-HEP}-91/68\\ October 1991
\end{flushright}
\vspace{1cm}

\centerline{\Large \bf Duality, Quotients, and Currents}
\vskip 1in

\centerline{\large Martin Ro\v cek\footnote{
 Permanent address: ITP, SUNY at Stony Brook, Stony Brook NY
11794-3840. \\ Email: rocek@dirac.physics.sunysb.edu} and Erik
Verlinde\footnote{ Email: verlinde@iassns.bitnet}}
\vspace{1.5cm}

\centerline{\it School of Natural Sciences}
\centerline{\it Institute for Advanced Study}
\centerline{\it Princeton, NJ 08540}

\vspace{1.5cm}

\centerline{\bf Abstract}
\vspace{.2cm}

\noindent
\parbox{15cm}{We study the generalization of $R\to 1/R$ duality to
arbitrary conformally invariant sigma models with an isometry. We show
that any pair of dual sigma models can be represented as quotients of
a self-dual sigma model obtained by gauging different combinations of
chiral currents. This observation is used to clarify the
interpretation of the generalized duality as a symmetry of conformal
field theory. We extend these results to $N=2$ supersymmetric sigma
models.}
\vfill
\eject


\newsection{Introduction}

\noindent
It is known from prehistoric times that the vacuum Maxwell equations
are preserved under the interchange of the field strength
$F_{\m\n}=\pa_\m A_\n -\pa_\n A_\m$ with its dual $\fr 12
\e^{\m\n\r\s}F_{\r\s}$.  This duality transformation exchanges the
electric and magnetic fields, and at the same time, the field
equations $\pa^\m F_{\m\n}=0$ with the Bianchi identities
$\e^{\m\n\r\s}\pa_\n F_{\r\s}=0$. In two dimensions a similar duality
transformation can be defined to act on the ``field strength'' of a
scalar field $P_\m = \pa_\m \f$:
\be
P_\m \leftrightarrow \e^{\m\n}P_\n \, .
\ee
Again, this exchanges the field equations $\pa^\m P_\m = 0$ with the
Bianchi identities $\e^{\m\n}\pa_\m P_\n = 0$.

More generally, in any two-dimensional nonlinear sigma model with an
isometry, one may define a duality transformation as follows: One
gauges the isometry, and adds to the action a Lagrange multiplier that
constrains the gauge field to be flat, \ie constrains its field
strength to vanish. This generalizes the well-known $R\ra 1/R$ duality
\cite{rdual}. In this paper we clarify the interpretation of this generalized
duality in terms of conformal field theory.  In particular, we show
that any dual pair can be obtained by gauging different combinations
of chiral currents in a higher dimensional sigma model, and extend
these results to $N=2$ superspace.

The plan of the paper is as follows: First we review duality and
discuss when it is an exact equivalence. In section 3 we discuss
isometries and the geometric significance of chiral currents. Our main
result is derived in subsection 3.2: we interpret duality in terms
conformal field theory quotients by chiral currents. This is used to
complete the discussion of duality as a symmetry of conformal field
theories. We end section 3 with some general remarks on duality. In
section 4, we extend the discussion to $N=2$ supersymmetric sigma
models. After reviewing superspace kinematics, in subsection 4.2 we
discuss axial and vector gauge fields and explain how they are used to
implement $N=2$ quotients and duality transformations. We then review
the explicit form of duality in $N=2$ superspace, and present the
construction of dual pairs as quotients.  Finally, we show how $N=2$
duality works in components.  In an appendix, we have added a brief
description of the geometry of $N=2$ quotients.

\newsection{Duality in non-linear sigma models}

\noindent
We consider the sigma model action
\be
S=\ind (g_{ij}+b_{ij})\pa x^i \pab x^j \, ,
\label{ssigma}
\ee
where $g$ is the metric tensor on some manifold and $b$ is a potential
for the torsion 3-form $T=\fr32db$. This action is invariant under the
isometry
\begin{equation}
\label{isometry}
\delta x^i=\epsilon\k^i
\end{equation}
when the vector field $\k^i$ satisfies Killing's equation ${\cal L}_\k
g_{ij}= \k_{i;j}+\k_{j;i}=0$ and in addition ${\cal L}_\k T = 0$; the
latter implies that locally
\begin{equation}
\label{Lie}
{\cal L}_\k b=d\omega.
\end{equation}
for some 1-form $\omega$. Here ${\cal L}_\k$ and $d$ denote the
space-time Lie- and exterior derivative. For sigma models with a
dilaton term
\be
S_{dil}=\ind \Phi R^{(2)}
\ee
the dilaton field $\Phi$ must satisfy $\k^i\Phi_{,i}=0$.  One can now
choose coordinates $\{x^i\}=\{ x^0,x^a\}$ such that the isometry acts
by translation of $x^0\equiv\th$, and all fields $g$, $b$ and $\Phi$
are independent of $\th$.  (Here we used the fact that the 2-form $b$
is defined only up to (space-time) gauge transformations $b\ra b +
d\l$, for some 1-form $\l$; the 1-form $\omega$ defined in (\ref{Lie})
transforms under this gauge symmetry as $\omega\to\omega+{\cal
L}_\k\l$.)

The dual theory can be found from the first order action
$S_1$\cite{sie,fj,b87}
\be
\eqalign{
S_1=\ind \Bigl[ g_{00}\A \AB & + (g_{0a}+b_{0a}) \A\pab x^a +
(g_{a0}+b_{a0})\pa x^a\AB\cr &+(g_{ab}+b_{ab})\pa x^a \pab x^b
+\td(\pa\AB -\pab\A )\Bigr]
\label{s1}\cr}
\ee
If one integrates out the Lagrange multiplier field $\td$, on a
topologically trivial worldsheet the gauge fields are pure gauge, $
\A = \pa \th$, $\AB = \pab \th$, and
one recovers the original model (\ref{ssigma}).  Classically, if one
integrates out the gauge fields $\A , \AB$ one finds the {\it dual\/}
model; this is a {\it new\/} theory with action $\du S$\cite{b87,b88}
\be
\du S = \ind (\du g_{ij} +\du b_{ij})\pa \xd^i \pab \xd^j \, ,
\label{sdual}
\ee
where $\{\xd^i\} = \{\td , x^a\}$ and
\be
\label{dual}
\eqalign{
&\du g_{00} = \fr1{g_{00}} \q \du g_{0a} = \fr{b_{0a}}{g_{00}} \q
\du g_{ab} = g_{ab} - \fr{g_{a0}g_{0b}+b_{a0}b_{0b}}{g_{00}} \q
\cr\cr
&\du b_{0a} = \fr{g_{0a}}{g_{00}} \q
\du b_{ab} = b_{ab} + \fr{g_{a0}b_{0b}+b_{a0}g_{0b}}{g_{00}} \, .
\cr}
\ee
Quantum mechanically, the duality transformation receives corrections
from the jacobian that comes from integrating out the gauge fields; at
one loop, this leads to a simple shift in the dilaton
\cite{b85,b87,b88}:\footnote{For some recent results on higher order
effects, see \cite{tse}}
\be
\label{dilshift}
\Phi \ra \Phi + \ln g_{00} \, .
\ee
With this shift, if the original theory was conformally invariant, at
least to one-loop order, the dual theory will be as well.  Note that
in general, the geometry of the sigma model is completely changed by
the duality transformation.

It is natural to ask if the dual models are truly equivalent as
conformal field theories.  We first consider global issues of the
procedure.  Let us assume that the isometry corresponds to a compact
$U(1)$-group, so that the coordinate $\theta$ is periodic with period
$2\pi$.  The constraint on $\A ,\AB$ that comes from integrating out
the Lagrange multiplier $\td$ implies $\A,\AB$ are flat, but in
principle they still may have nontrivial holonomies around
non-contractible loops.  In general, this would mean that the dual
theory will have twisted sectors, and the duality transformation
should then be thought of as an orbifold construction.\footnote{We
thank E. Witten for pointing this out to us.} However, by giving $\td$
the appropriate periodicity, the holonomies are also constrained to
vanish: the resulting winding modes of $\td$ act as Lagrange
multipliers for the holonomies. For example, for a worldsheet with the
topology of a torus the action (\ref{s1}) contains the term
\be
S_{wind}=n_a\!\oint_b\!A+n_b\!\oint_a\!A
\ee
where $n_a$ and $n_b$ are the winding numbers of $\td$ around the $a$-
and $b$-cycle on the torus. The sum over these winding numbers
produces a periodic delta-function for the holonomy, which by tuning
the periodicity makes the holonomy trivial as an element of $U(1)$.
Explicitly, we find that $\td$ should also have period $2\pi$.
Notice that this is consistent with the fact that for a $U(1)$ gauge
field the integrated curvature $\int\! F$ takes only integer values:
the integral over the periodic constant mode of $\td$ precisely yields
the required Kronecker delta (not a Dirac delta) to put $\int\! F$ to
zero. Thus we may conclude that the action (\ref{s1}) indeed is
equivalent to the original model without the gauge field.

\newsection{Duality and Quotients}

\noindent
The second step in the duality transformation is integrating out the
gauge field; this does not have an obvious interpretation in terms of
conformal field theory because the current associated with the
isometry in general does not correspond to a chiral current,
\ie to a Ka\v c-Moody type current algebra.  In this section,
we show how the general (abelian) duality transformations can be
reformulated as a construction that does involve chiral currents.
This construction will allow us to complete the proof that duality is
a true symmetry of conformal field theory.

\newsubsection{Isometries and chiral currents}

\noindent
As a preparation we investigate the following question: when do
isometries lead to chiral currents? We find that the Noether current
associated with the isometry (\ref{isometry}) is, in general
coordinates, given by
\ba
\label{currents}
\J\is \k^i(g_{ij}-b_{ij})\partial x^j+\omega_j \partial x^j\nonu
\JB\is \k^i(g_{ij}+b_{ij})\overline{\partial} x^j
-\omega_j\overline{\partial}x^j
\ea
where $\omega$ is the 1-form that appeared in (\ref{Lie}).  It is
easily shown using the equations of motion that this current is
conserved: $\partial \JB+{\bar \partial}\J=0$.  However, in general
the chiral components $\J$ and $\JB$ are not conserved separately, as
is illustrated by the familiar example of the Lorentz generators.

To have a chiral current $\J$, satisfying $\pab \J=0$, we must be able
to find a gauge for $b$ such that $\JB=0$. From (\ref{currents}) we
see that this implies that $\omega_j =\k^i(g_{ij}+b_{ij})$.  Inserting
this into (\ref{Lie}) gives a more restrictive condition on the
$b$-field, which combined with Killing's equation implies that the
vector $\k^i$ must be covariantly constant with respect to the
connection $D_i^+$
\be
\label{covconst}
D_i^+\k_j=\partial_i\k_j-\Gamma^k_{ij}\k_k-T^k_{ij}\k_k=0
\ee
where $T$ is the torsion tensor. At this point there are two
possibilities: $$
\mbox{(i) $\ T_{ij}^k\k_k=0\ , \qquad $ or $\qquad$
(ii) $\ T_{ij}^k\k_k\neq 0$.} $$ In the first case the condition
(\ref{covconst}) implies that the coordinate $\theta$ corresponding to
the Killing vector represents a free scalar that decouples from the
rest of the theory. Possibility (ii) is more interesting and leads to
an action of the form
\be
S_{L}=\ind \[\pa \th \pab \th +G_a(x)
\pa x^a\pab\th] \, +\,
S[x] \, ,
\label{Schir}
\ee
Clearly this action gives rise to a (left) chiral current
$\J^L=\partial\theta+\hf G_a\partial x^a$ that generates the chiral
symmetry $\delta \theta=\alpha(z)$.

A somewhat curious property of
this type of action is that by applying the duality transformation
(\ref{dual}) we get back the {\it same} action, \ie
\be
\label{sd}
g_{ij}=\du g_{ij}\, ,\qquad b_{ij}=\du b_{ij}.
\ee
In this sense a sigma model with a chiral current can be called
self-dual,\footnote{To really make it self-dual one should adjust the
periodicity of $\th$ to its self-dual value.} although one can
consider more general self-dual models for which (\ref{sd}) holds up
to diffeomorphisms.

\newsubsection{Duality and quotients by chiral currents}

\noindent
Even though the current that generates the isometry is in general not
chiral, the corresponding duality transformation can be understood in
terms of chiral currents.  The basic clue comes from a interesting
recent example: the stringy two dimensional black hole
\cite{wbl} constructed as a quotient of the $Sl(2,\R )$
WZW model by a $U(1)$ subgroup.  In \cite{dvv,giv,kir}, it was shown
that two different quotients (resulting from axial \vs vector
gaugings) gave equivalent CFT's with geometries dual in the sense
discussed above.  We now generalize this construction to
any pair of dual manifolds with dimension $d$.

We start with a $d+1$ dimensional sigma model which in addition to the
chiral current $\J^L$ has a right-handed chiral current $\JB^R$. From
the above discussion we see that this requires a second isometry with
a covariantly constant Killing vector with respect to the connection
$D_i^-$, which differs from $D_i^+$ in the sign of the torsion term.
The most general action then takes the form\footnote
{The action for the $SL(2,\R)$ WZW-model is of the form (\ref{Sgen})
with $\gr = \gl = 0$, $B(x)=\cosh (x)$, and $S[x] = -\ind\pa x\pab
x$.}
\be
S_{LR}=\ind \[\pa \th_L \pab \th_L + \pa \th_R \pab \th_R + 2 B(x)
\pa \th_R\pab \th_L +\gr\pa\th_R\pab x^a+\gl\pa x^a\pab\th_L\] \, +\,
S[x] \, ,
\label{Sgen}
\ee
where $x^a$ are coordinates on an arbitrary manifold with an arbitrary
WZW term, \ie $S[x]=\ind(G_{ab}+B_{ab})\pa x^a \pab x^b$.
This action has separately conserved left and right handed
currents
\be
\J^L = \j \, \, ; \qquad \JB^R=\jb \, ,
\ee
 corresponding to the $U(1)_R\times U(1)_L$ symmetry
$\d\th_R= \a_R({\bar z})\, ,\ \d\th_L=\a_L(z)$.
Next we can choose to gauge an axial or vector subgroup
\be
\d_{V,A} \th_R = \a \q \d_{V,A} \th_L = \pm \a \, .
\ee
For the gauged action we take the ``minimally coupled'' action that
follows from (\ref{Sgen}) by the prescription
\be
\label{mincoup}
\eqalign{
 \pa \th_{R} \ra \pa \th_{R} + \fr12\A \q
&\pa\th_{L} \ra
\pa\th_{L} \pm \fr12\A\, ,\cr
 \pab \th_{R} \ra \pab \th_{R} + \fr12\AB \q
&\pab\th_{L} \ra
\pab\th_{L} \pm \fr12\AB\, ,\cr}
\ee
and add the gauge invariant term
\be
\label{Sdiff}
\fr12(\th_R\mp\th_L)(\pa\AB -\pab\A ) \, .
\ee
We include this term so that the gauge-field couples directly to
chiral currents $\J^L$ and $\JB^R$ in the gauged action:
\be
\label{Sgau}
S_{gauged}^\pm =S_{LR}+\ind\[\A \JB^R \pm\AB \J^L + \fr12\A\AB (1\pm
B)\]\, .
\ee
This ensures that when the sigma-model described
by (\ref{Sgen}) is conformally invariant, the same holds for the
gauged models (and vice versa).  We come back to this point below.

Eliminating the gauge field from (\ref{Sgau}) gives
\be
\label{SV}
S_V=\ind\[ \Bigl(\fr{1+B}{1-B}\Bigr)\pa\td\pab\td
+\fr\gr{1-B}\pa\td\pab x^a +\fr\gl{1-B}\pa x^a\pab\td
+\fr12\fr{G^L_aG^R_b}{1-B}\pa x^a\pab x^b\] \, + \, S[x]
\ee
for the vector
gauging,
and
\be
S_A=\ind\[ \Bigl(\fr{1-B}{1+B}\Bigr)\pa\th\pab\th
+\fr\gr{1+B}\pa\th\pab x^a - \fr\gl{1+B}\pa x^a\pab\th
-\fr12\fr{G^L_aG^R_b}{1+B}\pa x^a\pab x^b\] \, + \, S[x]
\label{SA}
\ee
for the axial gauging, where
\be
\th = \th_R-\th_L\q\td =\th_R+\th_L\, .
\ee
We see that these actions describe different sigma models, each with
an isometry corresponding to translations in $\theta$ and $\td$
respectively.  If we choose
\be
\fr{1+B}{1-B}=g_{00} \q \fr{G^{R,L}_a}{1-B}=g_{0a}\pm b_{0a}\q
G_{ab}+B_{ab}+\fr12\fr{G^L_aG^R_b}{1-B}=g_{ab}+b_{ab} \, ,
\ee
we find that (\ref{SV}) and (\ref{SA}) are precisely related via
the duality transformation (\ref{dual}).   By taking into
account the jacobian that arises from integrating out the gauge field,
we find that the dilaton field gets shifted by
\be
\Phi\ra\Phi+\ln(1\pm B)\, ,
\ee
which is also in accordance with (\ref{dilshift}). It follows that {\it any}
pair of dual sigma models can be obtained in this way.

As mentioned above, the action (\ref{Sgen}) with chiral currents is
conformally invariant if the dual pair (\ref{SV}-\ref{SA}) are.  One
can explicitly check that this is true at one-loop order.  This is
most easily done using the previous observation that duality respects
the one-loop beta functions. We start with the gauged action
(\ref{Sgau}) and add a Lagrange multiplier term
\be
\ind\f(\pa\AB\! -\!\pab\A )\, .
\ee
Note that for duality transformations, it doesn't matter whether we
use ``minimal coupling'' (\ref{mincoup}) or the gauging (\ref{Sgau}),
as terms such as (\ref{Sdiff}) can be absorbed into the Lagrange
multiplier term.  We find that the dual of (\ref{Sgau}) with respect
to the vector (axial) symmetry is (after shifting
$\theta\ra\theta+\phi$) precisely the action $S_{V}$ ($S_A$) of
(\ref{SV},\ref{SA}), respectively, with an extra, {\it decoupled\/}
free field represented by $\phi$. This makes clear that when the
one-loop $\beta$-functions vanish for one of the three actions $S_A$,
$S_V$ and $S_{LR}$, the same holds for the other two.

We can now use the above quotient construction to fully establish that
the dual sigma models are equivalent as conformal field theories as
follows:  First we note that the two actions (\ref{SV}) and (\ref{SA})
are simply related by reversing the sign of $B$ and $G^L_a$. This
operation is a symmetry of the original action (\ref{Sgen}) in the
sense that it can be undone by the coordinate transformation
$\th_L\ra-\th_L$. The net effect of the combined operation, however,
is that the vector and axial gaugings are interchanged. From this we
may conclude that these two gaugings are, at least locally,
equivalent. To extend this to a global equivalence we need to impose
that $\th=\th_R-\th_L$ and $\td=\th_R+\th_L$ both have the same
periodicity.

Let us explain how all this can be interpreted in terms of conformal
field theory. The gauging of the currents amounts to a generalization
of the GKO coset-construction obtained by modding out by the left and
right chiral current. Concretely this implies that one only keeps the
fields which are primary with respect to the $U(1)_R\times U(1)_L$
current algebra.  The difference between the axial and vector gauging
is only in the assignment of the $U(1)$ quantum numbers
$(q_R,q_L)$ that are carried by these primary fields. Following the
above discussion we see that by reversing the signs of, say, all left
charges $q_L$, while keeping the same right charges $q_R$, we go from one
model to its dual. More precisely, for every primary field with charge
$(q_R,q_L)$ there is a corresponding field in the dual model with
charges $(q_R,-q_L)$.  All correlation functions are invariant under
this operation, because conformal dimensions, OPE's, etc.\ depend only
on quadratic combinations of the charges $q_L$. Thus we conclude that
duality is truly a symmetry of the conformal field theory.

\newsubsection{Some General Remarks on Duality}

\noindent
One of the most interesting applications of the quotient construction
and generalized duality is to the 2d black hole CFT.  For this we
refer the reader to the literature\cite{wbl,dvv,giv,kir}.  It is not
hard to think of other examples, but this we also leave to the reader.
Here we will just comment on some general aspects of duality.

A first remark concerns the occurrence of singularities. When the
isometry on the original manifold has a fixed point then by duality
the dual manifold will have a singularity at the corresponding point.
This can be illustrated with the somewhat trivial example of the 2d
plane.  In polar coordinates the metric
\be
ds^2=dr^2+r^2 d\theta^2
\ee
becomes
\be
ds^2=dr^2+r^{-2} d\theta^2
\ee
after duality. In addition there is a dilaton field $\Phi=\ln{r^2}$.
So indeed the fixed point of the isometry, \ie $r=0$, becomes a singular
point after the duality transformation.  The primary fields of the CFT
on the 2d plane all correspond to momentum modes. On the dual plane, however,
there are, due to the singularity, no (normalizable) momentum modes;
instead there are only winding modes. Duality tells us that, when we
identify the winding numbers $m$ with the angular momentum quantum
numbers $m$, corresponding dual fields have the same correlation
functions. Thus, although the dual theory appears to be a singular
theory, it actually behaves as the free theory on the plane.

For isometries corresponding to a compact $U(1)$, we can fix
the periodicity of $\th$ and $\td$ to be $2\pi$; different periods can
be obtained by rescaling $g_{00}$ and the other components of the
metric and $b$-field. Notice, however, that the period of $\td$ must
always be the reciprocal of that of $\th$; otherwise, the
two models are not fully equivalent but related via an orbifold
construction. As a limiting case we can consider the duality transformation
(\ref{dual}) for a non-compact isometry for which the corresponding
$\th$ coordinate can take any real value. We then find that the
dual coordinate $\td$ must have period zero, and so the
dual manifold is actually an orbifold obtained by modding
out the translations in $\td$. For the confromal field theory this implies
that the pure momentum states are removed and are again
replaced by winding states.

Duality can also be applied to models with a lorentzian signature.
  Particularly interesting examples are
backgrounds with a time-like killing vector $\k$, such as
static black hole solutions. Here it may happen that in some region of
space-time $\k$ becomes space-like. This means that at the boundary
between the two regions, usually called the event horizon, $\k$ is a
null vector. Since we have the relation
\be
g_{00}=\|\k \|^2
\ee
we see that, as for fixed points, there will be a
singularity in the dual space-time. Thus, the observation made in
\cite{giv,dvv} that the event horizon and the singularity of the 2d
black hole are interchanged by duality, is in fact much more general.
Finally, we note that in principle the duality transformation (\ref{dual})
could change the signature of the manifold. This would be quite disturbing,
but fortunately one can check that
\be
\det \du g_{ij}={1\over g_{00}^2}\det g_{ij},
\ee
which shows that at least determinant of the metric does not change
sign. We believe therefore that the signature of space-time is not
altered by duality.

\newsection{Duality in $N=2$ Supersymmetric Sigma Models}

\noindent
Somewhat surprisingly, the whole previous discussion can be lifted to
$N=2$ superspace. First we introduce some notation.

\newsubsection{N=2 Superspace}

\noindent
In $N=2$ superspace, dynamics and kinematics are entwined in
interesting ways.  In addition to the choice of action, one has to
choose representations of supersymmetry, \ie superfields that obey
various kinematical constraints (see for example \cite{book}).  Here
we focus on two kinds of multiplets that are naturally dual to each
other: chiral and twisted chiral multiplets \cite{ghr}.  (Other
representations are possible \cite{blr}).  The basic derivatives of $N=2$
superspace are complex left(right) moving spinor derivatives
$D_+, D_-$ that obey the algebra
\be
\{D_+ ,\bd_+\}=\pa \q \{D_- ,\bd_-\} = \pab \, ,
\label{eq19}
\ee
all other (anti)commutators vanishing.  The $N=2$ superspace action is
written as\cite{ghr}
\be
S=\ins K(\F , \FB , \L , \LB ) \, ,
\label{eq20}
\ee
where $\F$ and $\L$ are complex superfields obeying a chiral or
twisted chiral constraint:
\be
\bd_\pm \F =0\q\bd_+\L = D_-\L = 0\, .
\label{eq21}
\ee
As a consequence of these constraints, the action changes only by a
total derivative if we shift $K$ by functions $f(\F ,\L )$, $g(\F ,\LB
)$, or the complex conjugates.  The constraints can be explicitly
solved by writing $\F,\L$ in terms of an unconstrained complex
superfield $\Psi$:
\be
\F=i\bd_+\bd_-\Psi\q \L=i\bd_+D_-\Psi \, .
\label{eq22}
\ee
If $K$ is independent of the twisted chiral fields $\L , \LB$, then
the manifold described is K\"ahler\cite{zum}; in the general case, the
geometry has been described in \cite{ghr}.

\newsubsection{N=2 Quotients and Duality}

\noindent
Because the geometry is restricted by $N=2$ supersymmetry, one can not
take a usual quotient.  In the K\"ahler case, the situation is well
known \cite{bw,hitch,hull}, and one must perform a K\"ahler reduction.
We describe the construction for the general case (\ref{eq20}) in the
appendix.

In practice, the $N=2$ quotient is performed very much in analogy with
the bosonic quotient; we simply gauge the symmetry, {\it not\/}
including a kinetic term for the gauge field, and eliminate the gauge
field by its resulting equation of motion.  Supersymmetry guarantees
that the resulting quotient includes the moment map constraint
described in the appendix.  To do this, we need to review the
superfield description of gauge fields, and introduce a (slight)
modification of the known gauge multiplet, the twisted gauge multiplet
that couples naturally to twisted chiral superfields.\footnote{One can
also find ``chiral'' gauge multiplets that couple to the semichiral
models of \cite{blr} in an obviously analogous way.  Chris Hull has
informed us that he has independently observed that such variant gauge
multiplets can exist in two dimensions.} The usual gauge multiplet
naturally gauges a vector symmetry, whereas the twisted multiplet
naturally gauges an axial gauge symmetry.  Both the usual gauge
multiplet and the twisted gauge multiplet are described by a real
unconstrained superfield $V$. They differ in that for the usual gauge
multiplet the gauge transformation acts as a shift by the real part
of a chiral superfield,
\be
\d V = \F + \FB \, ,
\label{eq23}
\ee
whereas for the twisted multiplet the gauge transformation acts as a
shift by the real part of a {\it twisted\/} chiral superfield:
\be
\d V= \L + \LB \, .
\label{eq24}
\ee
The corresponding super field-strengths also differ: for the usual
gauge multiplet, the complex twisted chiral scalar $\bd_+D_-V$ is
gauge invariant, whereas for the twisted multiplet, the field-strength
is chiral: $\bd_+\bd_-V$. (These field-strengths correspond to the
physical scalars of the two-dimensional gauge multiplet). The only
isometries that we can gauge without losing $N=2$ supersymmetry are
ones that act holomorphically either on chiral superfields or on
twisted chiral superfields.  For the $U(1)$ case, the situation is
particularly simple, and one can always choose coordinates such that
$K$ is invariant and depends only on $\F +\FB$ or $\L + \LB$,
depending on what kind of a gauging we want to perform.  The
``minimal'' coupling prescription consists of merely substituting
\be
\F +\FB\ra \F +\FB +V
\label{eq241}
\ee
or
\be
\L +\LB\ra \L +\LB + V\, .
\label{eq242}
\ee
However, that is not the whole story. There exist further gauge
invariant terms (generically referred to as ``Fayet-Iliopoulos
terms''\cite{fi,bw}) that can and should be added.  The original FI
term consisted of just $V$; because the gauge parameters are
constrained, up to a total superspace derivative, this is gauge
invariant.  In our case, when we have both chiral fields ($\F$) and
twisted chiral fields ($\L$), we can generalize this; for an arbitrary
function $f$
\be
[f(\L ) + \bar f(\LB )]V
\label{eq243}
\ee
is invariant (again, modulo total superspace derivatives) under chiral
gauge transformations (\ref{eq23}), and
\be
[f(\F ) + \bar f(\FB )]V
\label{eq244}
\ee
is invariant under twisted chiral transformations (\ref{eq24}).
However, just as in the bosonic case, when we perform a duality
transformation these terms can be absorbed by shifting the Lagrange
multipliers: The condition that $V$ is pure gauge is imposed by
constraining the appropriate field strengths, \ie by including a term
\be
i(\Psi \bd_+D_- V + \bar\Psi D_+\bd_- V)
\label{eq245}
\ee
for the usual gauge multiplet (\ref{eq23}), or
\be
i(\Psi \bd_+\bd_- V + \bar\Psi D_+D_- V)
\label{eq246}
\ee
for the twisted gauge multiplet (\ref{eq24}).  Here $\Psi$ is an
unconstrained Lagrange multiplier field.  Integrating by parts, these
become
\be
(\L +\LB )V \q (\F + \FB )V \, ,
\label{eq247}
\ee
respectively, where $\L$,$\F$ are now constrained Lagrange
multipliers.  Clearly, (\ref{eq247}) can be used to absorb
(\ref{eq243},\ref{eq244}).

\newsubsection{Explicit superspace duality}

\noindent
Concretely, consider a model with a superspace Lagrangian \cite{ghr}
\be
K=K(\F + \FB , X^a) \, ,
\label{eq25}
\ee
where $X^a$ are ``spectator'' superfields that may be of any type
(chiral, twisted chiral, etc.).  The first order Lagrangian is
\be
K_1=K(\F + \FB +V,X^a) -(\L +\LB )V \, .
\label{eq26}
\ee
Integrating out $\L$ constrains $V$ to be pure gauge, and hence leads
back to (\ref{eq25}); integrating out $V$ gives
\be
\du K = K(V,X^a)-(\L +\LB )V
\label{eq27}
\ee
where
\be
\fr{\pa K}{\pa V} = \L + \LB
\label{eq28}
\ee
defines $V=V(\L +\LB ,X^a)$.  Thus $\du K$ is the Legendre transform
of $K$\cite{lr}.

This discussion has been purely classical; quantum mechanically, one
generally needs a dilaton as in the bosonic case.  Curiously, in some
circumstances (for example when the dual pair actually has an
accidental $N=4$ supersymmetry), the dilaton contribution is
automatically incorporated \cite{b85,b87,b88}.

As in the bosonic case, we can learn about duality and conformal field
theory quotients from the two-dimensional black hole.  In
\cite{rss,arss}, the supersymmetric WZW model on $SU(2)\times U(1)$
was described in $N=2$ superspace in terms of one chiral and one
twisted chiral superfield, and was shown to be dual to
$[SU(2)/U(1)]\times U(1)^2$.

We can follow the basic pattern and generalize this to arbitrary $N=2$
models with an appropriate isometry (as described above).  We take a
superspace Lagrangian\footnote{ For $SU(2)\times U(1)$\cite{rss,arss},
$$
K=
\int^{\textstyle\fr{\F\FB}{\L\LB}}
\, \fr{dx}{x}\,\ln (1+x)\, -\,\fr12(\ln\L +\ln\LB )^2 \, .
$$
where $\ln \L$ plays the role of $\L$ in the text here, etc.  Up to
total superspace derivatives, this can also be written as
$$ K= -
\int^{\textstyle\fr{\L\LB}{\F\FB}}
\, \fr{dx}{x}\,\ln (1+x) \, +\,\fr12(\ln\F +\ln\FB )^2\, ,
$$
These are the forms (\ref{eq38}) and (\ref{eq37}) of the superspace
action specialized to $SU(2)\times U(1)$.}
\be
K=-\fr12(\L +\LB )^2 +K_0(\F+\FB+\L+\LB ;X^a) \, ,
\label{eq38}
\ee
where, as above, $X^a$ are spectator fields.  Up to a total superspace
derivative, this can be rewritten as
\be
K=\fr12(\F +\FB )^2 +\hat K_0(\F+\FB+\L+\LB ;X^a) \, ,
\label{eq37}
\ee
where
\be
\hat K_0( \F+\FB+\L+\LB ; X)= K_0 -\fr12(\F+\FB+\L+\LB )^2 \, .
\label{eq39}
\ee
As we saw above, quotients are inherently ambiguous; it is much safer
to make a duality transformation.  As in the bosonic case, we will
find that after a duality transformation, we find the quotient model
and a completely decoupled free field.  We can make a duality
transformation either with respect to $\F +\FB$ or $\L + \LB$; we use
(\ref{eq38}) or (\ref{eq37}) as our starting point.  Performing the
duality transformation as described above (dropping total superspace
derivatives), for the vector case we find
\be
\eqalign{
K^+&=-\fr12(\L +\LB )^2 +K_0(V;X)-(\du\L + \du\LB )(V-\F -\FB -
\L - \LB ) \cr
&=-\fr12(\hat\L +\hat\LB )^2 +\fr12(\du\L +\du\LB )^2 +K_0(V;X) -
(\du\L +\du\LB )V \, ,}
\label{eq40}
\ee
where $\hat\L = \L -\du\L$ is a free field (and is decoupled), and $V$
is determined by
\be
\fr{\pa K^+}{\pa V}=\fr{\pa K_0}{\pa V} - \du\L -\du\LB = 0 \, .
\label{eq41}
\ee
For the axial gauging, we find
\be
\eqalign{
K^-&=\fr12(\F +\FB )^2 +\hat K_0(V;X)+(\du\F + \du\FB )(V-
\F -\FB -\L - \LB ) \cr
&=\fr12(\hat\F +\hat\FB )^2 +K_0(V;X) -\fr12(V-\du\F -\du\FB )^2
\, ,}
\label{eq42}
\ee
where $\hat\F = \F -\du\F$ and $V$ is determined by
\be
\fr{\pa K^-}{\pa V}=\fr{\pa K_0}{\pa V}-V +\du\F + \du\FB = 0 \, .
\label{eq43}
\ee
After dropping the free field terms, the superspace actions $K^+(\du\L
+\du\LB ;X)$ (\ref{eq40}) and $K^-(\du\F +\du\FB ;X)$ (\ref{eq42}) are
indeed Legendre transforms of each other, and hence correspond to dual
models.  To show that any dual pair can be derived in this way, we
note (\ref{eq40}) is a free field action for $\hat\L$ combined with
$\du K_0(\du\L + \du\LB ;X ) +\fr12(\du\L +\du\LB )^2 $, where $\du
K_0$ is the Legendre transform of $K_0$.  Hence if we are given some
$K(\du\L + \du\LB ;X)$, and want to find $K_0(V;X)$, we simply take
the inverse Legendre transform of $K(\du\L + \du\LB ;X)-\fr12(\du\L +
\du\LB )^2$.

\newsubsection{$N=2$ Duality in Components}

\noindent
Though the $N=2$ construction is in many ways analogous to the bosonic
construction, the relation can be made more precise by working out the
component form. As the basic superspace technology is readily
available (see for example \cite{book,ghr,hitch}), we give only the
relevant results.  In particular, since we are interested in comparing
to the bosonic expressions of the previous sections, we drop all
fermionic terms.

To proceed, we must specify the constraints on the spectator fields
$X^a$; we choose to divide them into two sets, chiral superfields
$X^i$ and twisted chiral superfields $X^r$.  We further introduce the
following notation:
\be
\rho = \re (\F +\L ) \sq \ff = \re (\F -\L ) \sq \th_R = \im
(\F +\L ) \sq \th_L = \im (\F -\L ) \, ,
\ee
for the leading bosonic components of the superfields $\F , \L$, and
\be
x^a = \re X^a \q y^a = \im X^a
\ee
for the leading bosonic components of the ``spectator fields'' $X$.
It is also useful to express the answer not in terms of $K_0$ or $\hat
K_0$ above (see (\ref{eq39}), but in terms of the average $\bk =\hf
(K_0 +\hat K_0)$.

The bosonic component action that results from (\ref{eq38},\ref{eq37})
is then
\be
\eqalign{
S=\ind \[ &\pa \th_L \pab \th_L + \pa \th_R \pab \th_R +
4\bk''\pab\th_L \pa \th_R +\gr\pa\th_R\pab x^a-
\gl\pa x^a\pab\th_L\cr
&+\pa (\ff + \bk' )\pab (\ff + \bk' )\]
\, +\, S[\rho ,x,y] \, ,
}
\label{compS}
\ee
where $\bk'\equiv\fr{\pa \bk (\rho ; x,y)}{\pa \rho}$, etc., and
\be
\begin{array}{rclrclrclrcl}
G^R_{x^i}\is 2\bk'_{x^i}&G^R_{x^r}\is -2\bk'_{x^r}& G^R_{y^i}\is
-2\bk'_{y^i}&G^R_{y^r}\is 2\bk'_{y^r}\\[3mm] G^L_{x^i}\is
2\bk'_{x^i}&G^L_{x^r}\is 2\bk'_{x^r}& G^L_{y^i}\is
-2\bk'_{y^i}&G^L_{y^r}\is -2\bk'_{y^r}
\end{array}
\ee
This is precisely of the form (\ref{Sgen}) with an extra free field
$\ff + \bk'$.  This free field that is present before duality explains
why $N=2$ duality decouples a {\it complex} field.

\vspace{1cm}
\noindent
{\sc Acknowledgements:} We would like to thank P. Ginsparg, A. Giveon,
E. Witten for discussions and M. Best for help with \TeX. EV wishes
to thank the Aspen Center for Physics, where part of this work was done.
The work of MR is supported in part by NSF grant No. PHY 91-08054 and
by the John Simon Guggenheim Foundation, and that of EV by the W.M. Keck
Foundation.

\vspace{1cm}
\noindent{\sc Appendix: The geometry of $N=2$ supersymmetric quotients}

\def\I{{\cal J}}
\def\Li{{\cal L}}

\noindent
We begin by reviewing the geometry of $N=2$ models as described in
\cite{ghr}.  In general, $N=2$ supersymmetric models can be
constructed on complex manifolds with two complex structures $\I^\pm$
that are both compatible with the metric $g$, and that are preserved
by connections $\Gamma^\pm =\Gamma^0\pm T$: $(\pa +\Gamma^\pm )\I^\pm
=0$.  Here $\Gamma^0$ is the Levi-Civita connection constructed just
out of the metric and $T$ is the torsion $T=\fr32db$ as in the text. It
follows that $\o^\pm_{ij}\equiv g_{ik}\I^{\pm k}{}_j$ is in general
not closed but obeys $\pa^\pm \pab^\pm \o^\pm=0$.  If the torsion
vanishes, then $\I^+=\I^-$, the manifold is K\"ahler and $d\o =0$.
Quotients will preserve $N=2$ supersymmetry if and only if they
preserve these geometric properties.

In the K\"ahler case, the $N=2$ quotient can be performed for any
isometry that preserves the K\"ahler form.  The procedure is well
known: since $\o$ is a (closed) symplectic form, the equation
$\Li_\k\o=0$ can be integrated to $$
\o_{ij}\k^j=\pa_i\mu \, ,
$$ which defines the {\it moment map} $\mu$ up to an integration
constant.  The K\"ahler quotient is then simply the ordinary quotient
on the submanifold $\mu = 0$.

A generalization to the more general $N=2$ geometry exists when the
following conditions are satisfied (we have not determined which of
them are independent and which are implied by a subset): There must be
a family of submanifolds on which the commutator $[\I^+,\I^-]$ of the
two complex structures vanishes. Each such submanifold further admits
two families of smaller submanifolds on which $\I^+=\pm\I^-$; any of
these smallest submanifolds is {\it K\"ahler}, where the K\"ahler form
is the restriction of $\o^\pm$.

The isometry must preserve both complex structures as well as the
torsion, \ie $\Li_\k\I^\pm=\Li_\k T = 0$. Furthermore, the isometry
must be tangent to one of the two families defined above; without loss
of generality, we assume $$
\I^{+i}{}_j\k^j =\I^{-i}{}_j\k^j \, .
$$ We now construct the moment map $\mu$ on this K\"ahler submanifold.
The integration constants are allowed to depend on the coordinates of
the other family of submanifolds as long as $$
(\pa^+-\pa^-)(\pab^+-\pab^-)\mu = 0 \, .  $$ The quotient is again
taken on the submanifold $\mu =0$.  This generalizes the usual moment
map construction.

\font\twelvebf=cmbx12
\newcommand{\savebold}{\fam\bffam\twelvebf}
{\renewcommand{\Large}{\normalsize}
\renewcommand{\bf}{\sc}
\newcommand{\np}{Nucl.\ Phys.\ }
\newcommand{\pr}{Phys.\ Rev.\ }
\newcommand{\cmp}{Commun.\ Math.\ Phys.\ }
\newcommand{\pl}{Phys.\ Lett.\ }

}

\end{document}